\documentclass[journal=jacsat,manuscript=article]{achemso}
\usepackage[version=3]{mhchem} 

\newcommand\mb{\mathbf}
\usepackage{amssymb}
\usepackage{appendix}

\author{Zuzanna M. Jedlinska}
\affiliation{Department of Physics and Astronomy, University of Pennsylvania, Philadelphia, United States}

\author{Robert A. Riggleman}
\affiliation{Department of Chemical and Biomolecular Engineering, University of Pennsylvania, Philadelphia, United States}
\email{rrig@seas.upenn.edu}

\title{The effects of dynamic binding on the phase behaviour and properties of polymer blends undergoing complex coacervation}

\abbreviations{}

\begin{document}


\begin{abstract}

Associative polymer networks have shown a major promise in fabrication of self-healing and responsive materials. The can also serve as simple models to study more complex biological systems where transient interactions play an important role. In this work we investigate the properties of charged polymer blends whose constituents are capable of creating dynamic bonds. We model dynamic bonds as harmonic springs with additional bond formation energy, $\varepsilon_a$, which can be adjusted to influence the reaction rates for binding and unbinding. We show that varying the number of binding sites on the chains and $\varepsilon_a$ has a major effect on the resulting phase diagram. We further investigate the diffusive behaviour of the coacervates. We also study, how the network structure changes with varying number of active site, and increasing $\varepsilon_a$, and identify the values which result in the network percolation. Lastly, we explore the possibility of inducing orthogonal phase separation by by introducing a second type of binding site, which cannot interact with the first kind.

\end{abstract}

\section{Introduction}

Dynamic binding has emerged as an attractive way to control self-assembly and macrophase separation in polymeric materials \cite{mester_macro-_2011, Zheng_dynmaic_covalent, h_bonding} where a relatively low concentration of binding monomers dispersed along a polymer backbone can lead to drastic property changes. As opposed to static polymer networks, dynamic binding allows for the properties and topology of these materials to be tuned using external stimuli \cite{lee_phase_2007, bottle_brush, elliott_supramolecular_2009, dynamic_networks}. Dynamic polymer networks have a vast array of applications in material science ranging from self-healing rubbers and shape memory materials \cite{lewis_review_2016, sani_intrinsic_2022, Zheng_dynmaic_covalent, shape_mem_review} to hydrogels \cite{hydrogel} and wearable electronics \cite{wearable_electronics}. An important feature of dynamic polymer networks is their reusability and being relatively easy to recycle. The energy of dynamic bonds is usually lower than that for permanent covalent linkages, and thus the associations are reversible using only relatively mild elevation in temperature \cite{Tang2018, Chino2001ThemoreversibleCR, recycling}. This reduces the cost of reprocessing and also ensures that the properties of the constituent chains are not degraded during the recycling process. 
In biomedical applications, dynamic polymer networks can serve as a novel platform for drug delivery, gene therapies, and act as biosensors \cite{biomedical, genes}.
\textit{In vivo}, reversible associations can drive the formation of membraneless organelles and facilitate chromatin condensation through transient protein-DNA associations \cite{shorter_phase_2019, uversky_intrinsically_2017, gomes_molecular_2019}. In general, however, systems where the phase behaviour is influenced by dynamic bonding processes remain poorly understood for a simplified, molecular modeling perspective.

Dynamic binding itself can be realized in multiple ways, for example via hydrogen or sulfur bonding \cite{h_bonding, sulfur_dynamic}, through the use of a metal cross-linker \cite{metal}, or by joining the chains through reversible covalent bonds \cite{covalent_bonds, covalent}. Thus, the exact mechanism in which the network is created can be chosen to suit a particular application. To be able to capture this complicated phenomenon, computational and theoretical methods have been used to create a guiding framework \cite{fredrickson_coherent_2018, mohan_field-theoretic_2010, gibbs, triblock, simulations} to study dynamical polymer network. In particular, field-theoretic approaches~\cite{fredrickson_2007} have been successful in elucidating the trends governing the behaviour of the systems capable of supramolecular self-assembly. However, these studies predict equilibrium structures, which assumes that experimental systems will be able to equilibrate on laboratory time scales. Depending on the strength of the dynamic bonds and the propensity for the system to form percolating networks, equilibrium may not be achievable. Once the details of the system are replaced with a simplified, coarse-grained model, one can understand which properties are general and emerge due to the dynamic binding itself, and which one are specific only to particular systems. However, the study of this important class of materials on mesoscopic length scales still remains a challenge.

In this study, we bridge this gap by performing coarse-grained simulations of dynamically bonded polymer networks in the mesoscopic range of system sizes. We are able to access this size regime through the use of efficient coarse-grained simulation method called Theoretically Informed Langevin Dynamics (TILD). This is a hybrid particle/field method, which leverages the advantages of both these approaches to reduce the computational cost associated with simulating systems with large amount of particles. We study how the dynamical and static properties of homogeneous polymer blends are affected by the number of active sites participating in dynamic binding, and the energy associated with bond formation.

Most of the previous theoretical and experimental work on associative polymer networks \cite{danielsen_phase_2023, danielsen_chemical_2023, sticker-clustering, stickers-high-dens} have focused on neutral or sparsely charged systems. Under those circumstances, the transient bonds were the main factor influencing the behaviour and determining the properties of the studied systems. In this work, however, we focus on the effect of dynamic binding treated only as a small perturbation on top of a dominant electrostatic force rather than as a dominant factor. 
Since associative interactions can in principle be controlled in an orthogonal manner to the electrostatic forces, by selecting the species with desired chemical affinity, dynamic bonds can provide an additional way to fine-tune the properties of the condensates. While there have been extensive computational and theoretical studies of the phase behaviour of polymer solutions driven solely by coacervation \cite{chang_sequence_2017,Delaney,riggs} or dynamic bonding \cite{Rubinstein-gelation, danielsen_chemical_2023, viscoelastic, dynamics-of-networks, cross-linked}, the behaviour of systems with multiple driving forces for phase separation have received less attention.

In addition, our study can help to elucidate the details of essential biological processes.
Simulations in which molecules can create transient bonds have already been used to understand the origin and the nature of self-assembly occurring within the living cells \cite{mitra_coarse-grained_2023}, and to study cytoskeletal dynamics \cite{freedman_versatile_2017}.
Another important biological process, whose driving forces can be captured in our simulations, is liquid-liquid phase separation (LLPS). The significance of this type of phase separation stems from the fact that during LLPS both the polymer-rich and the polymer-lean phases maintain their liquid-like properties. \textit{In vivo}, many biomolecules interact though non-selective, weak attractive forces, such as hydrogen bonding. Yet, these biomolecules are still able to precisely phase-separate into specific intracellular compartments. Although still not fully understood, this phenomenon has been demonstrated experimentally and through computational modelling to depend on the presence of specific binding motifs present on distinct species \cite{bio-sep, welles_determinants_2023}. We believe that our system can serve as a model to investigate the role of these motifs on LLPS propensity. 

The outline of the paper is as follows: In Section II we describe the TILD method, and provide the details about the simulation setup and the parameter choices. A detailed description of the computational algorithms can be found in the Appendix. Then, in Section III we present the results. We begin with the analysis of the bulk properties. Next, we study the dynamical properties of the coacervates, and analyze the connectivity and the topology of the emergent networks. Finally, we investigate the possibility of orthogonal phase separation when we introduce another type of active sites into the system. We close the paper with a summary and possible future directions in Section IV.

\section{Methods}
\label{sec:methods}

\subsection{Polymer Model and Theoretically Informed Langevin Dynamics (TILD)}

\begin{figure}
    \centering
    \includegraphics[width=0.75\textwidth]{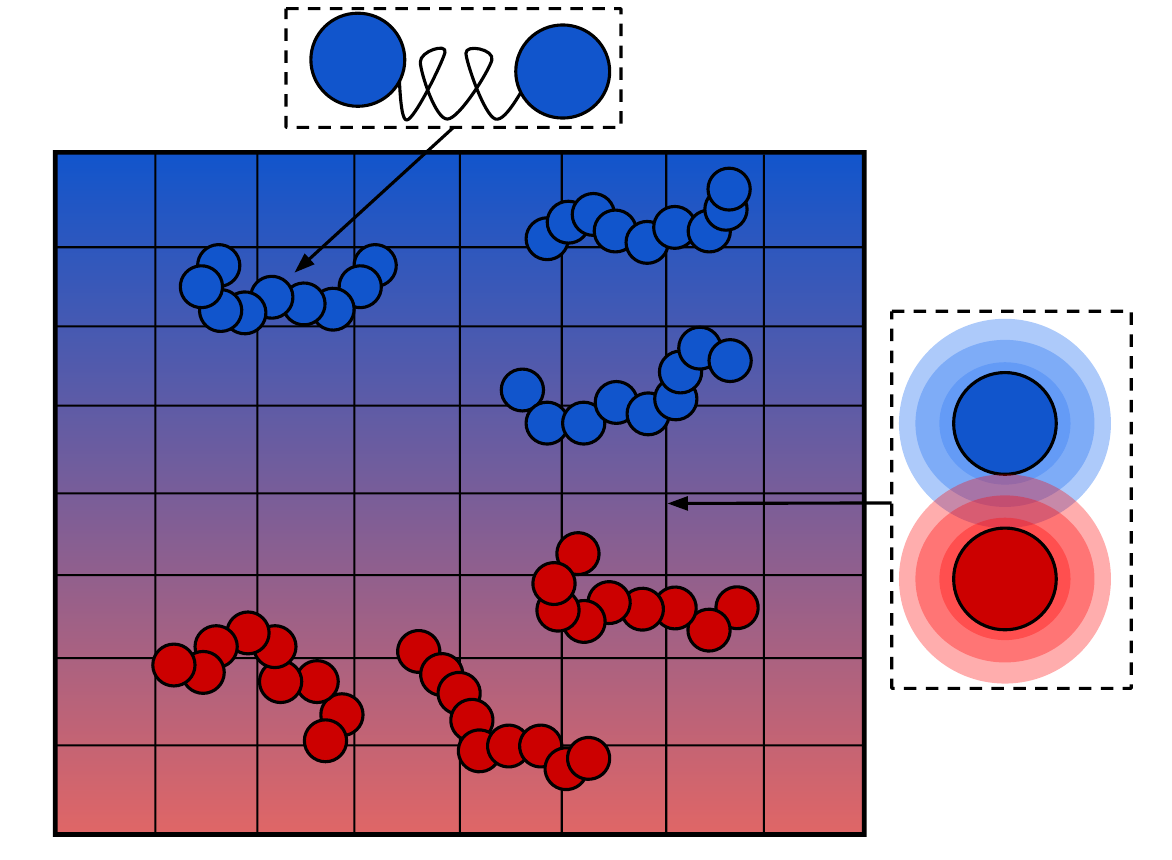}
    \caption{Schematic representation of the TILD method. Explicit coordinates are used in order to calculate bonded interactions, whereas non-bonded bonded interactions use density-field representation instead.}
    \label{fig:tild}
\end{figure}

We begin by describing the details of our simulation approach. We employ a hybrid particle/field method called Theoretically Informed Langevin Dynamics (TILD). The fields used correspond to the mass or charge density and are obtained from discretizing particle densities over the grid using a particle-to-mesh scheme similar to those used in Ewald summations {\cite{deserno1998mesh}}. The density fields are implemented as a computational device to efficiently calculate non-bonded interactions present in the system by avoiding using explicit coordinates, while bonded interactions are still computed using explicit particle coordinates. A schematic representation of the TILD approach is shown in Figure~\ref{fig:tild}.
All simulations presented in this manuscript were performed using MATILDA.FT, our in-house GPU-accelerated software which is available open-source on GitHub. A detailed discussion of the TILD method, the associated theoretical background, and the implementation details can be found in our previous manuscript announcing the public release of the code \cite{jedlinska_matildaft_2023}. 

We model polymers as discrete Gaussian chains \cite{Gaussian_chain} with the bonded potential given by
\begin{equation}
    \label{eq:bonds}
    \beta U_0 = \sum_j^n \sum_s^{N-1} \frac{3}{2b^2} |\mathbf{r}_{j,s} - \mathbf r_{j,s+1}|^2,
\end{equation}
where $b$ is the statistical segment length, $\mathbf{r}$ is a vector denoting the position of the monomer. The variables $j$ and $s$ denote the particular chain index and the monomer index along the chain, respectively.

We use an implicit solvent model, where the solvent-mediated forces between the monomers are taken through Edwards potential \cite{edwards, fredrickson_equilibrium_2006, villet2014efficient}, given by
\begin{equation}
    \beta U_1 = \frac{u_0}{2} \int d\mathbf{r} \int d\mathbf{r}' \, 
    \hat \rho(\mathbf r) 
    \, 
    u_G(|\mathbf r-\mathbf r'|)
    \,
    \hat \rho(\mathbf r').
    \label{eq:edwards}
\end{equation}
In Equation \ref{eq:edwards}, $\hat \rho_{A/B}$ is the grid-based density field of the monomers of type A or type B. The $u_0$ parameter is responsible for adjusting the strength of the excluded volume interactions between the monomers, and a positive value corresponds to the good solvent since there is net repulsion between the monomers. $u_G$ is a unit Gaussian potential, $u_G(r) = (2\pi\sigma^2)^{-\mathbb{D}/2} e^{-r^2/2\sigma^2}$, $\mathbb{D}$ is the dimensionality of the system, and $\sigma$ controls the range of the interactions. No distinction is made between the positively and negatively charged monomers in the excluded volume potential. Electrostatic interactions are resolved by numerically solving the Poisson equation for the electrostatic potential, $\phi(\mathbf r)$,
\begin{equation}
    \label{eq:poisson}
    \nabla^2 \phi(\mathbf r) = -4\pi l_B \breve \rho_c(\mathbf r),
\end{equation}
where $l_B$ is the Bjerrum length and $\breve \rho_c(\mathbf r)$ is the Gaussian-smeared charge density. The charge density is given by
\begin{align}
\breve \rho_c(\mathbf r) = \sum_j q_j \, u_C(\mb r-\mb r_j), \\
u_C(|\mb r|) = (2\pi\sigma_C^2)^{-\mathbb{D}/2} \; e^{-|\mb r|^2/2\sigma_C^2}.
\end{align}
Here $\sigma_C$ is the charge smearing length and can be different from the smearing length used for mass density.

All simulations are performed in an orthogonal 3-dimensional box with a $25 \times 25$ base and the length of 185 along the z-directions. All lengths are reported in the units of the statistical segment length of the polymers, denoted as $b$ and set to unity. The simulations are initialized from a dense ``slab'' which is placed in the center of the box, and the polymers are allowed to relax away from this dense state during the equilibration procedure. This allows us to efficiently generate a single, polymer-rich phase without the need to wait on coarsening dynamics of droplets formed randomly in our simulation box. Simulations last $10^8$ time steps, and the data is gathered after allowing the systems to equilibrate for $5 \cdot 10^7$ time steps. The values of the numerical parameters used in the simulations are listed for convenience in the Table \ref{tbl:parameters}.

\begin{table}
  \caption{Summary of the simulation parameters}
  \label{tbl:parameters}
  \begin{tabular}{ll}
    \hline
    Parameter & Value\\
    \hline
    N (chain length) & 75 \\
    $n$ (number of chains) & $1234$ \\
    $L_x \times L_y \times L_z$ (box lengths) & $25 \times 25 \times 185$\\
    $\varepsilon_a$ (bond formation energy) & $0-11$ \\
    $\mathcal{M}$ (number of active sites) & $\lbrack 2,3,4 \rbrack$ \\
    $r_0$ (dynamic bond equilibrium distance) & 0 \\
    $k_{spring}$ (dynamic bond spring constant) & 3\\
    \hline
  \end{tabular}
\end{table}

\subsection{Dynamic Binding}

\begin{figure}
    \centering
    \includegraphics[width=0.75\textwidth]{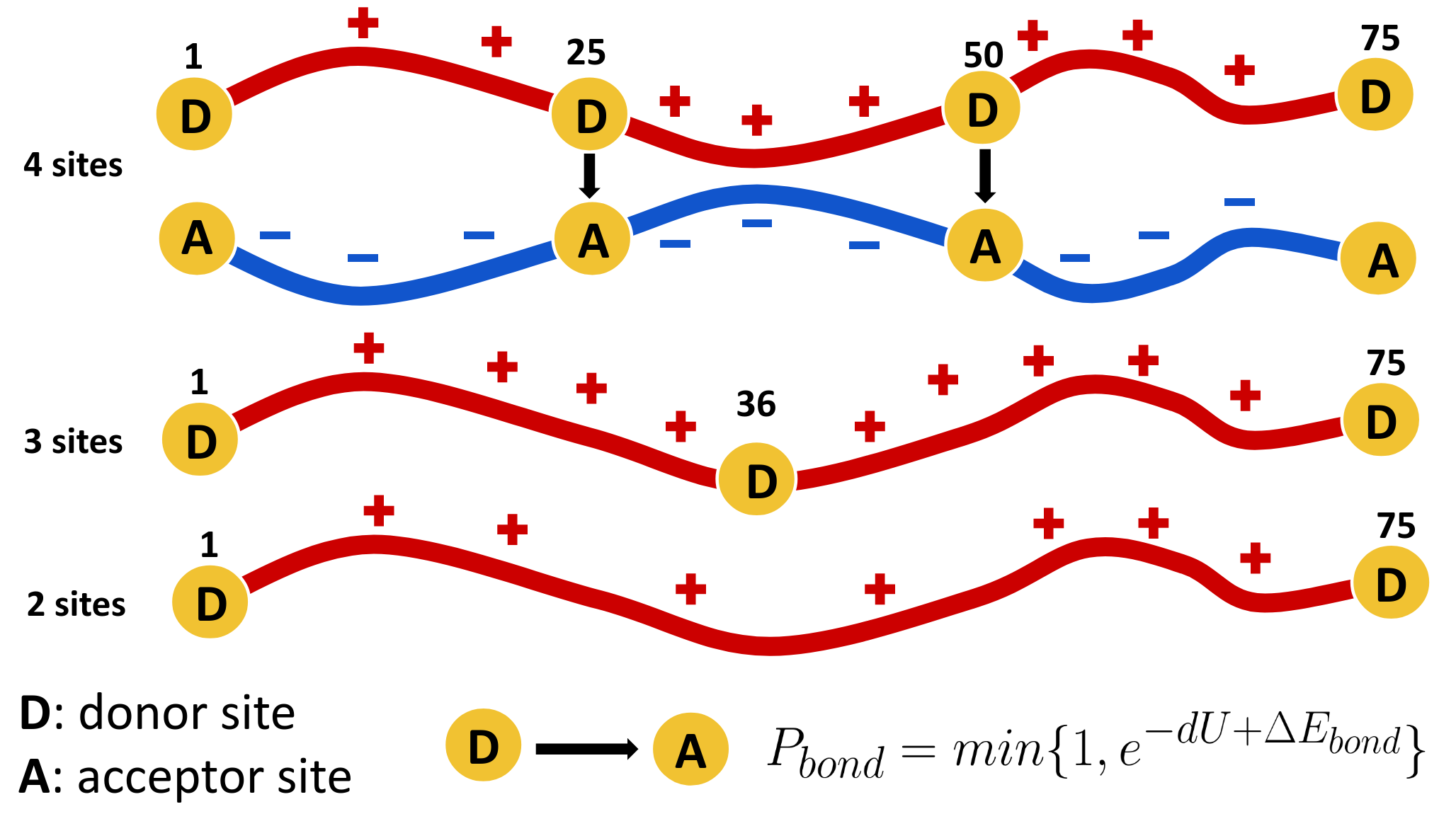}
    \caption{Schematic representation of the polymer chains with variable number of active sites. (Top): Schematic representation of a bonded pair of a donor (red) and acceptor (blue) chains. Numbers denote the position of the active sites along the chains with 4 (top), 3 (middle), and 2 (bottom) active sites.}
    \label{fig:model}
\end{figure}

All simulations consist of an equimolar blend of positively and negatively charged polymer chains in an implicit solvent. All chains have the total length of $N=75$ monomers and can carry 2, 3 or 4 active sites which participate in reversible binding. These active sites are distributed as illustrated in \ref{fig:model} and their count on a chain is denoted as $\mathcal{M}$. Two types of active sites, donor and acceptor sites, are present in the simulation. Each chain can only carry active sites belonging to one of these types. Donor sites are always placed on positively charged chains, whereas acceptor sites are placed on negatively charged chains. Thus, in our system we have positively charged donor chains and negatively charged acceptor chains.

We enforce a binary binding constraint such that any donor can only be attached to one acceptor at any given time. 
We model dynamic bonds as harmonic springs, governed by the same potential as given in eq.~\ref{eq:bonds}. The probability of bond creation, $P_{\mathrm{accept}}$ is taken through Metropolis acceptance criterion $P_{\mathrm{accept}} = \min[1, e^{(-\Delta E)}]$. Here, $\Delta E$ refers to change in energy upon the acceptance of the move. $\Delta E$ depends on the extension of the harmonic spring created or destroyed upon binding or unbinding, and on $\varepsilon_a$ which is additional energy which serves as a measure of affinity between the donors and acceptors. All energies are reported in the units of $k_BT$, where $k_B$ is the Boltzmann constant, and $T$ denotes the temperature. As in all of our calculations, we set $k_BT$ to be equal to unity. A bonding move occurs every 10,000 time steps during which the simulation is advanced in time using the TILD method. We verified that, as required for a system at equilibrium, the particular choice of the frequency of binding attempt did not change the thermodynamic quantities. The detailed description of of the algorithm and the parallel implementation can be found in Appendix A, and is omitted here for brevity.


\section{Results and discussion}
\label{sec:results}

In this section we present the results obtained from the simulations of homogeneous blends of charged polymers carrying variable number of active sites, $\mathcal{M}$. We investigate how the properties of the coacervates created by these chains change as a function of the bond energy, $\varepsilon_a$. 
All concentrations are reported as a dimensionless number, $C^{*}$, which denotes the monomer density multiplied by $\frac{R_g^3}{N}$. $R_g$ is the the radius of gyration of the ideal polymer chain, given by $R_g = \sqrt{\frac{N b^2}{6}}$ with $b$ as the statistical segment length and $N$ is the number of monomers of the polymer chain.

\subsection{Phase Diagram}
\label{sec:phase-diagram}

\begin{figure}
    \centering
    \includegraphics[width=0.75\textwidth]{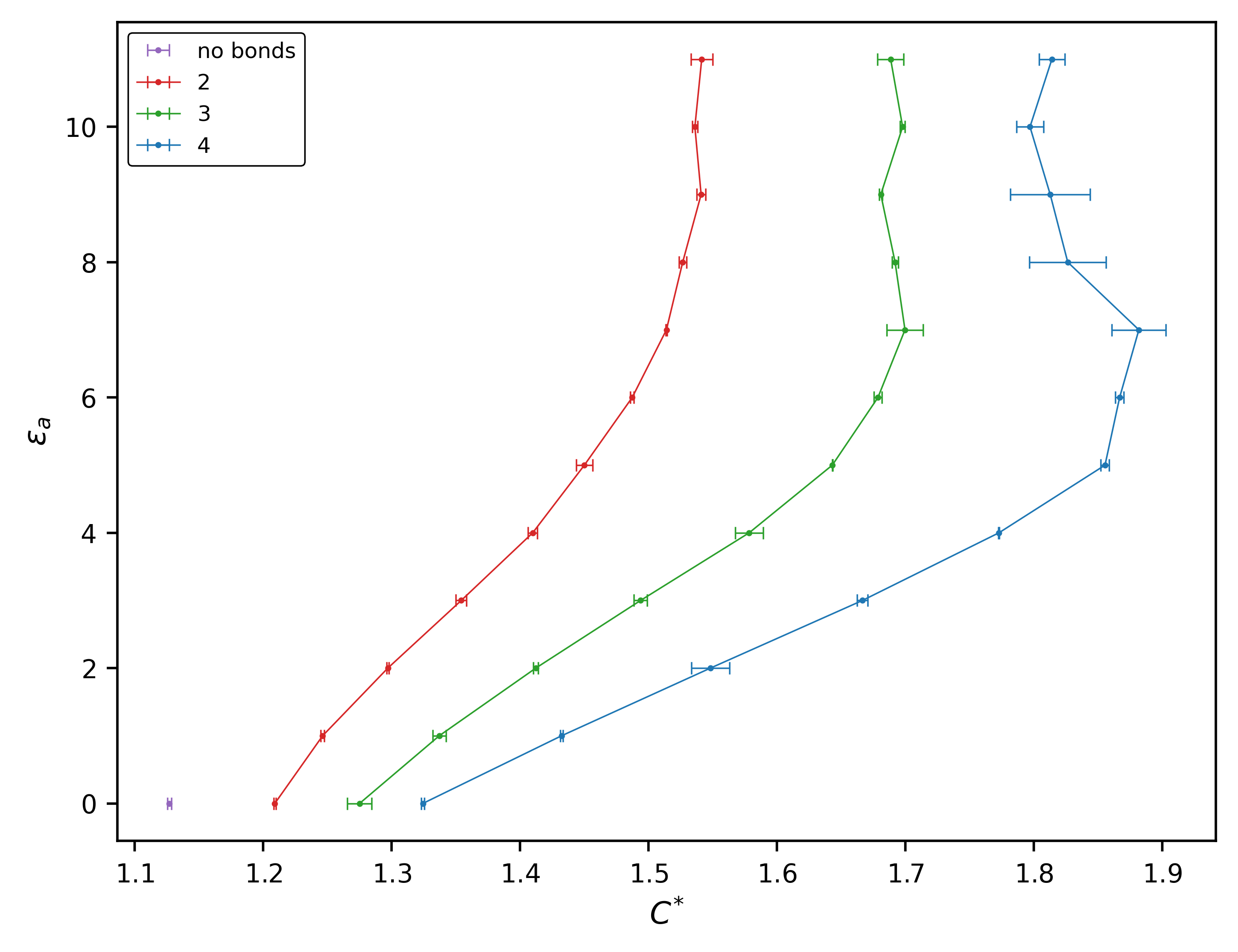}
    \caption{Phase diagram of systems with variable number of active sites per chain, $\mathcal{M}$. Data points corresponding to $\mathcal{M}=2$ are colored red, $\mathcal{M}=3$ are green, and $\mathcal{M}=4$ are blue. The purple data point corresponds to a system where dynamic bonds were turned off. The data points correspond to the fit to the density at the center of the coacervate. For clarity, we omit the dilute branch as the concentration there is invariantly 0 (none of the coacervates dissolve).
    At low values of $\varepsilon_a$ the density of the coacervates steadily increases and saturates at high values of $\varepsilon_a$ (in some cases paradoxically decreasing). This saturation is due to the kinetic arrest of the coacervates. It is clearly visible in the case of $\mathcal{M}=4$ where at the highers values of  $\varepsilon_a$ the coacervates are unable to equilibriate within the computationally accessible time.
    The error bars denote one standard deviation of the data between the independent simulation runs.}
    \label{fig:density-diagram}
\end{figure}

We begin by investigating how the density of coacervates changes as a function of the number of active sites on the chain, $\mathcal{M}$, and the affinity energy between the donor and acceptor sites, $\varepsilon_a$. The results for the chains with $\mathcal{M}=2,3$ and $4$ are plotted in Figure~\ref{fig:density-diagram}.
At low values of $\varepsilon_a$ the density of all coacervate steadily increases. This magnitude of this change increases with increasing value of $\mathcal{M}$. The density if the coacervate becomes higher due to the increasing amount of harmonic bonds which draw the particles closer together.
At intermediate and high values of $\varepsilon_a$, only small or none increase in the density is observed. This is due to the active sites becoming saturated or, at high energies, due to the coacervates becoming kinetically arrested and being unable to equilibriate to reach higher density.
At a fixed value of $\varepsilon_a$, chains with higher number of active sites always develop coacervates with higher polymer concentration, as they are able to create more harmonic bonds.
In the following sections, we analyze the the changes in the microscopic details of the coacervates and how they relate to the phase diagram.

\subsection{Bond Fraction and Persistence Time}

\begin{figure}
    \centering
    \includegraphics[width=1\textwidth]{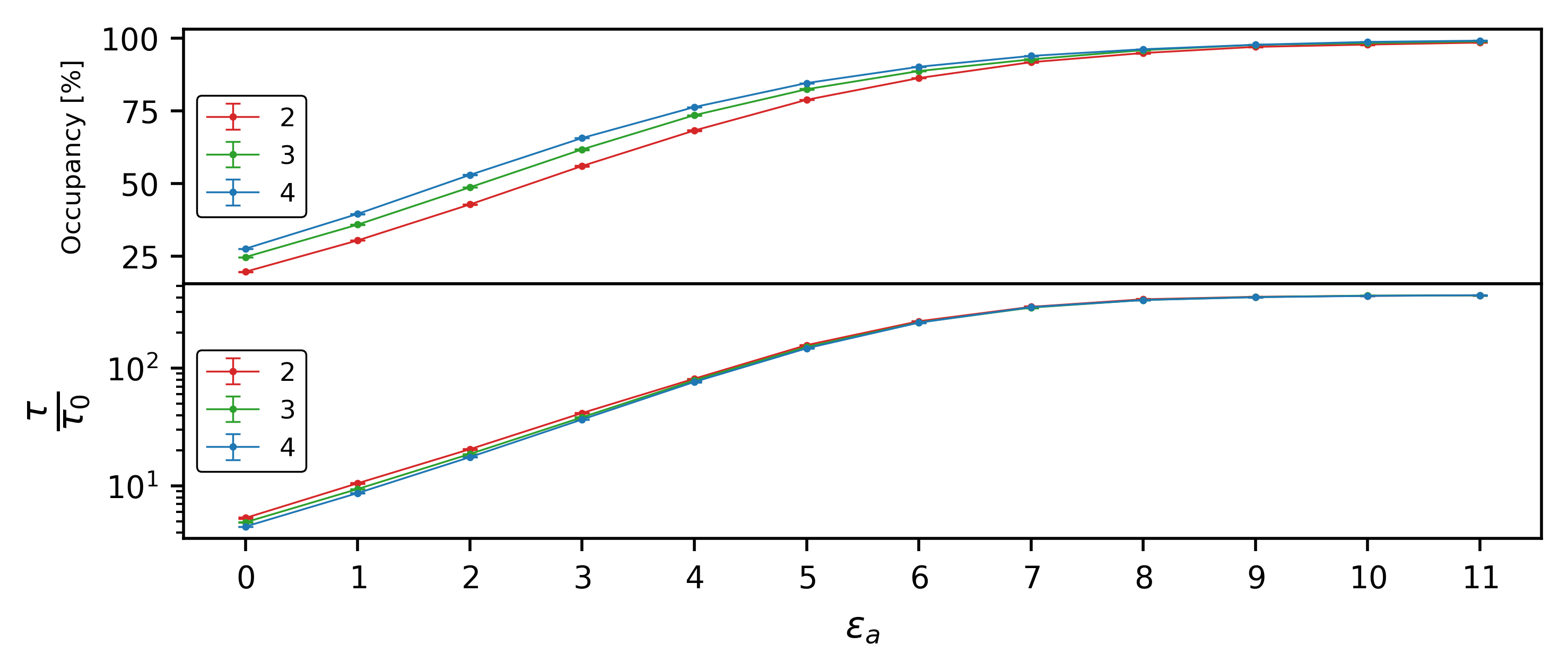}
    \caption{Occupancy of the active sites as a function of $\varepsilon_a$ (top) the mean bonds persistence time of the dynamic bonds (bottom) in the systems with $\mathcal{M}=2$ (red), 3 (green), and 4 (blue).
    (Top): At low energies exponential growth of occupancy is observed as a function of $\varepsilon_a$. The growth ceases at the saturation energy $\varepsilon_a \sim 8$, independent of the value of $\mathcal{M}$.
    (Bottom): Mean bond lifetime. The reported values are scaled by the time required for all active sites to participate in a MC step, $\tau_0 = {\tau_{MC}}{\mathrm{active fraction}}$, where active fraction refers to the fraction of particles active at each move, and $\tau_{MC}$ is the time interval between the subsequent MC moves. Initially, the bond lifetime increases exponentially. Past $\varepsilon_a \sim 8$ the bond lifetime approaches the measurement time $(300 \tau_{0})$}.
    \label{fig:bond-frac}
\end{figure}

After analyzing the bulk properties of the coacervates, we next focus on the structure of the underlying dynamic network. We begin by quantifying the occupancy of active sites as a function of $\varepsilon_a$ and $\mathcal{M}$. The results are displayed in the top panel of Figure~\ref{fig:bond-frac}. At the corresponding values of $\varepsilon_a)$ chains with more active sites achieve higher occupancy. At high values of $\varepsilon_a$ the active sites become saturated and the occupancy approaches $100\%$ for all values of $\mathcal{M}$.

Another important feature of the dynamical network is the mean lifetime of the dynamic bonds, which we plot as a function of $\varepsilon_a$ in the bottom panel of Figure~\ref{fig:bond-frac}. We calculate the bond lifetime by measuring the number of time steps between the formation of a given bond and the time when it breaks (or when the measurement time time is exceeded if the bond does not break) and taking the average over all bonds.
Similar to the active site occupancy, up until the saturation point, the lifetime of dynamic bonds scales increases with increasing $\varepsilon_a$. Specifically, it scales as $e^{(\mathcal{A} \varepsilon_a)}$, where $\alpha$ is an arbitrary constant, in agreement with \cite{danielsen_phase_2023} and \cite{yasuda_coarse-grained_2023}. The authors in \cite{yasuda_coarse-grained_2023} show that the value of $\mathcal{A}$ depends on the strength of the spring associated with the dynamic bond. In our case, we estimate $\mathcal{A}\sim \frac{2}{3}$.
Past the saturation point the chance of a bond breaking is low, thus the persistence time approaches the measurement time.
In contrast to the active site occupancy, only negligible variations are observed at the corresponding values of $\varepsilon_a$ for the chains with different $\mathcal{M}$.

\subsection{Mean-Squared Displacement and Dynamic Properties}
\label{sec:dynamics}

\begin{figure}
    \centering
    \includegraphics[width=1\textwidth]{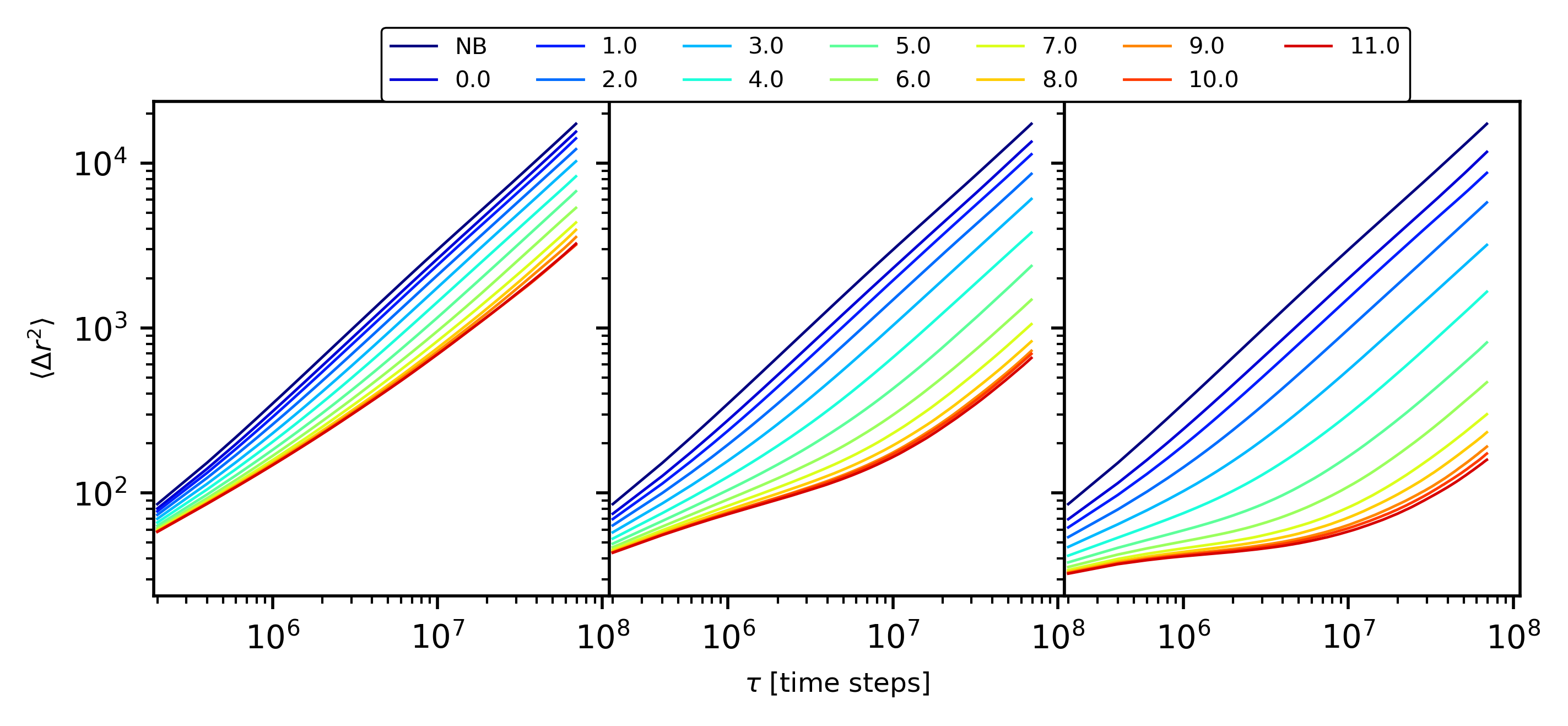}
    \caption{Mean-squared displacement ($\langle \Delta r^2 \rangle$) over time for the monomers belonging to chains with $\mathcal{M}=2$ (left), 3 (middle), and 4 (right). The data points labeled NB correspond to the control system with no dynamic bonds present.}
    \label{fig:MSD-log-log}
\end{figure}

Next, we focus on the dynamic properties of the coacervates, and investigate how these are affected by variations in the number of active sites, $\mathcal{M}$, and the value of $\varepsilon_a$. We begin by calculating the average mean-squared displacement, $\langle \Delta r^2(\tau) \rangle$, of the monomers for different systems. For brevity, we drop the explicit time dependence of $\Delta r^2(\tau)$ and refer to it simply as $\Delta r^2$. We plot $\langle \Delta r^2 \rangle$ as a function of time for system with varying $\mathcal{M}$ on log-log scale in Figure~\ref{fig:MSD-log-log}.

The value of $\langle \Delta r^2 \rangle$ increases according to the relationship $\langle \Delta r^2\rangle \sim \tau^{\alpha}$, where $\tau$ denotes time, and $\alpha$ is the power law scaling exponent.
Significant differences can be seen between the plots corresponding to different values of $\mathcal{M}$. The separation of the values of $\langle \Delta r^2\rangle$  between the consecutive bond energies increases with increasing $\mathcal{M}$. When $\mathcal{M}=2$, systems at and intermediate values of $\varepsilon_a$ display only Fickian behaviour $(\alpha =1)$. At the highest values of $\varepsilon_a$, initial subdiffusive $(\alpha < 1)$ behaviour is observed, which at long time-scales becomes Fickian diffusion.
When $\mathcal{M} = 3$ and 4, the subdiffusive region is more pronounced. In particular, when $\mathcal{M} = 4$ $\varepsilon_a \geq 8$ molecules display caging at short time-scales, where they are unable to dissociate from the network, and remain confined within a small volume. At longer time-scales, they are finally able to dissociate and diffuse from the original confinement.

The differences between the systems with different values of $\mathcal{M}$ are likely due to the cooperative effect of multiple binding sites on the chain. Specifically, when multiple binding sites are present, all of them need to associate in order for the chain to be freed from the network and diffuse. Since we assume that all binding and unbinding events are independent, the probability of the chain escaping the network is $\sim p_{\mathrm{off}^{\mathcal{M}}}$, where $p_{\mathrm{off}}$ is the probability of any given bond breaking.

\begin{figure}
    \centering
    \includegraphics[width=0.9\textwidth]{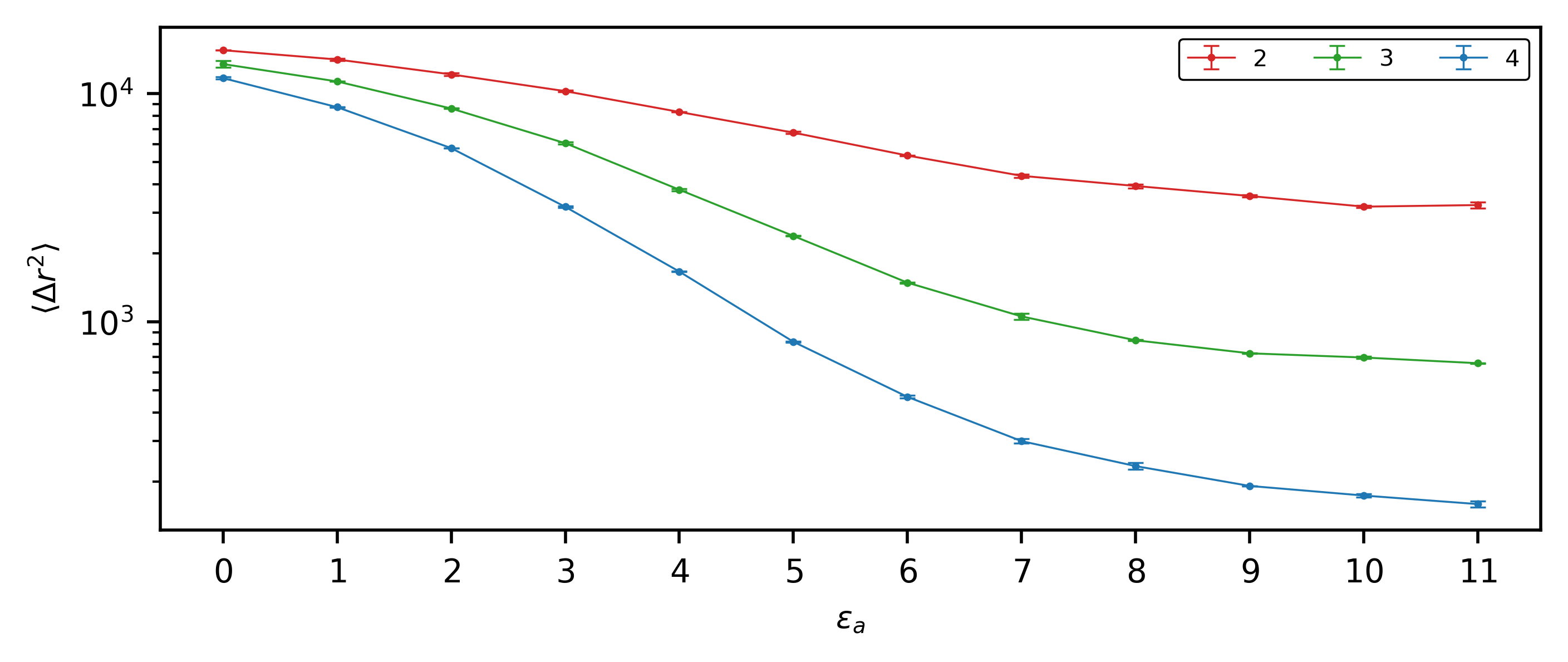}
    \caption{The terminal value of the mean-squared displacement ($\langle d^2 \rangle$) measured over an interval of $7 \cdot 10^7$ time steps as a function of energy for the chains with 2 (red), 3 (green), and 4 (blue) active sites. }
    \label{fig:MSD}
\end{figure}

For easier comparison, in Figure~\ref{fig:MSD}, we also plot the values of $\langle \Delta r^2\rangle$ after $10^8$ time steps for the systems with variable $\mathcal{M}$. The choice of $10^8$ time steps for the time scale is arbitrary, and our results are not qualitatively sensitive to this choice. As expected, for all values of $\mathcal{M}$ we observe that the magnitude of the mean-squared displacement decreases with increasing $\varepsilon_a$ due to the formation of a dynamic network and gelation.
At all the corresponding values of $\varepsilon_a$, systems with higher $\mathcal{M}$ have a lower value of the magnitude of $\langle \Delta r^2 \rangle$. This separation due to different values of $\mathcal{M}$ becomes more pronounced with increasing $\varepsilon_a$. The plateau in $\langle \Delta r^2 \rangle$ which sets in $\varepsilon_a \sim 8$ for all systems indicates that coacervates become kinetically arrested.

Previous experimental studies have shown that under certain conditions associative polymer networks can exhibit superdiffusive behaviour on short time-scales \cite{star-diffusion, stickers-high-dens}. The underlying mechanism for this apparent superdiffusion is ``hopping", where a chain completely dissociates from the network and diffuses freely over a large distance \cite{stickers-mechanism, multi-step}. This mechanism has been shown to be greatly facilitated by looping (not present in our system), where the chain binds to itself, reducing the number of free active sites, thus facilitating the hopping events. Experiments are usually carried at the concentrations much smaller than these used in our simulations.
Due to these differences, we do not observe superdiffusion in our systems. We expect that tuning both the density, and the reaction kinetics according to the experimental values would potentially allow us to reproduce the superdiffusive behaviour. However, that would likely results in systems that are either too large to simulate, or the required equilibration time of these systems would be to long to be handled by our current hardware resources (using slower reaction kinetics). Since diffusion is a dynamical property, and the focus of this manuscript is mainly on the thermodynamic properties (which insensitive to the reaction kinetics nor the diffusive behaviour) we do not investigate that matter further. Instead, we focus on a more in-depth characterization of the thermodynamic properties of the coacervates.

\begin{figure}
    \centering
    \includegraphics[width=0.75\textwidth]{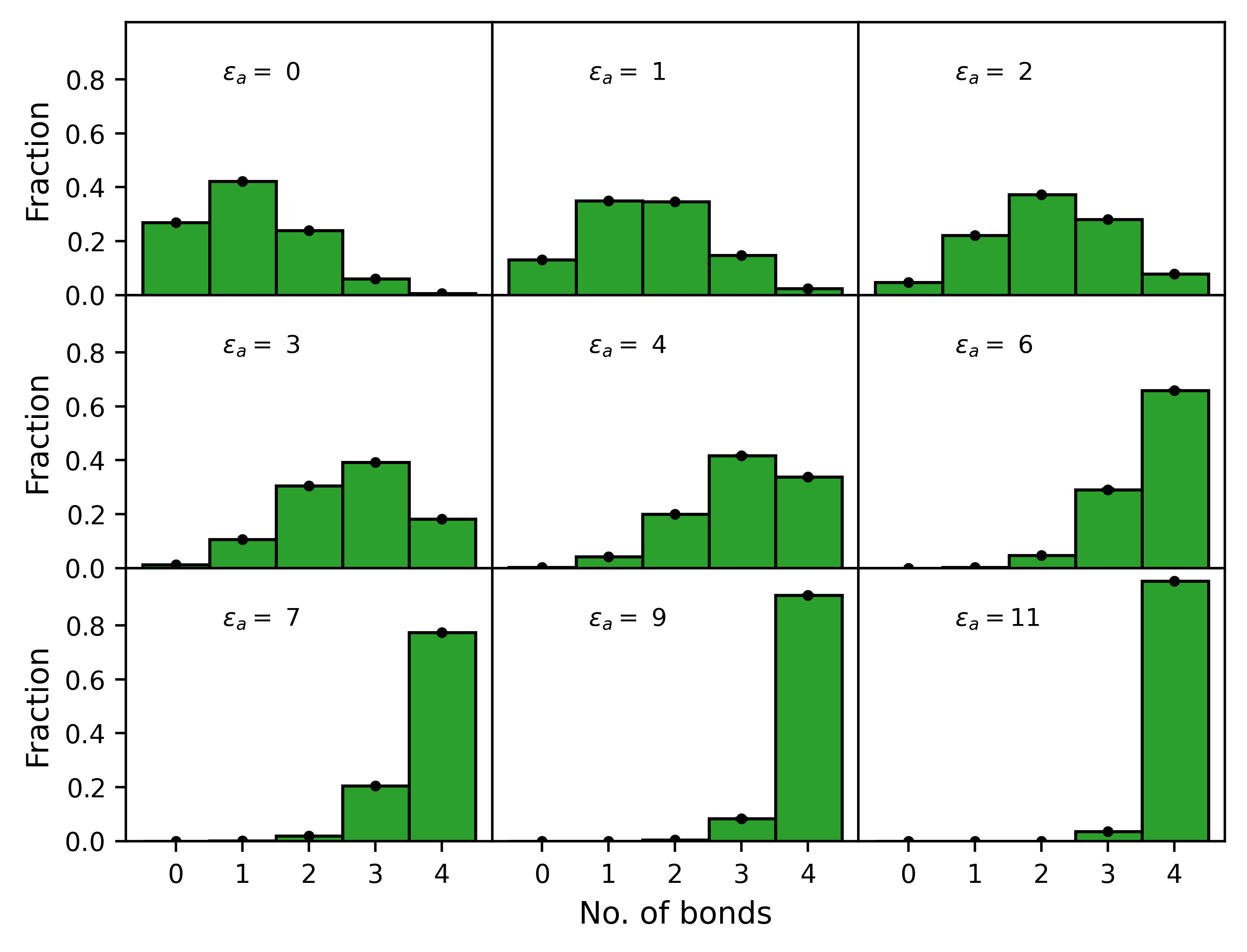}
    \caption{Histogram of the number of bonds created by chains with 4 active sites. Chains without any bonds correspond to the unbound ``hoppers''. Error bars, indicating one standard deviation, are smaller than the data points (black dots) on top of the histogram bars.}
    \label{fig:Bins}
\end{figure}

To investigate the link between the observed diffusive behaviour and the underlying, we investigate, how the occupancy fraction of a single chain changes as a function of $\varepsilon_a$. We focus on the systems with $\mathcal{M} = 4$ and display the results in Figure~\ref{fig:Bins} for selected values of the binding affinity energy. We observe that at low values of $\varepsilon_a$ there exists a significant population of fully unbound chains. As this population diminishes with increasing $\varepsilon_a$, the diffusive behaviour transitions from regular diffusion to the subdiffusive regime. Almost no free chains are observed at the highest values of $\varepsilon_a$,and most of the chains have their active sites fully saturated. Accordingly, for these systems we observed kinetic arrest and caging, as can be seen in the Figure~\ref{fig:MSD-log-log}.

\subsection{Reaction Kinetics}
\label{sec:kinetics}

In this section, we focus on the reaction kinetics, and how they relate to the behaviour observed on the phase-diagram, and to the diffusive properties.
We begin by measuring the rate constants, $k_{\mathrm{on}}$ and $k_{\mathrm{off}}$, associated with the conversion of the active sites between the free and occupied state. We assume our system can be described by an equilibrium chemical reaction where the state of the active site changes according to
\begin{equation}
 \ce{$\lbrack D_f \rbrack$ + $\lbrack A_f \rbrack$ <=>[\ce{k_{\mathrm{on}}} ][\ce{k_{\mathrm{off}}} ] $\lbrack A-D \rbrack$},
 \label{eq:reaction}
\end{equation}
where $\lbrack D_f \rbrack$ and $\lbrack A_f \rbrack$ refers to the concentration of the free (unreacted) donors or acceptor sites, respectively, and $\lbrack A-D \rbrack $ is the concentration of the bound pairs. The equilibrium constant is then given by the ratio of these rate constants, 
\begin{equation}
 K_{eq}=\frac{k_{\mathrm{on}}}{k_{\mathrm{off}}}.
 \label{eq:keq}
\end{equation}

\begin{figure}
    \centering
    \includegraphics[width=.75\textwidth]{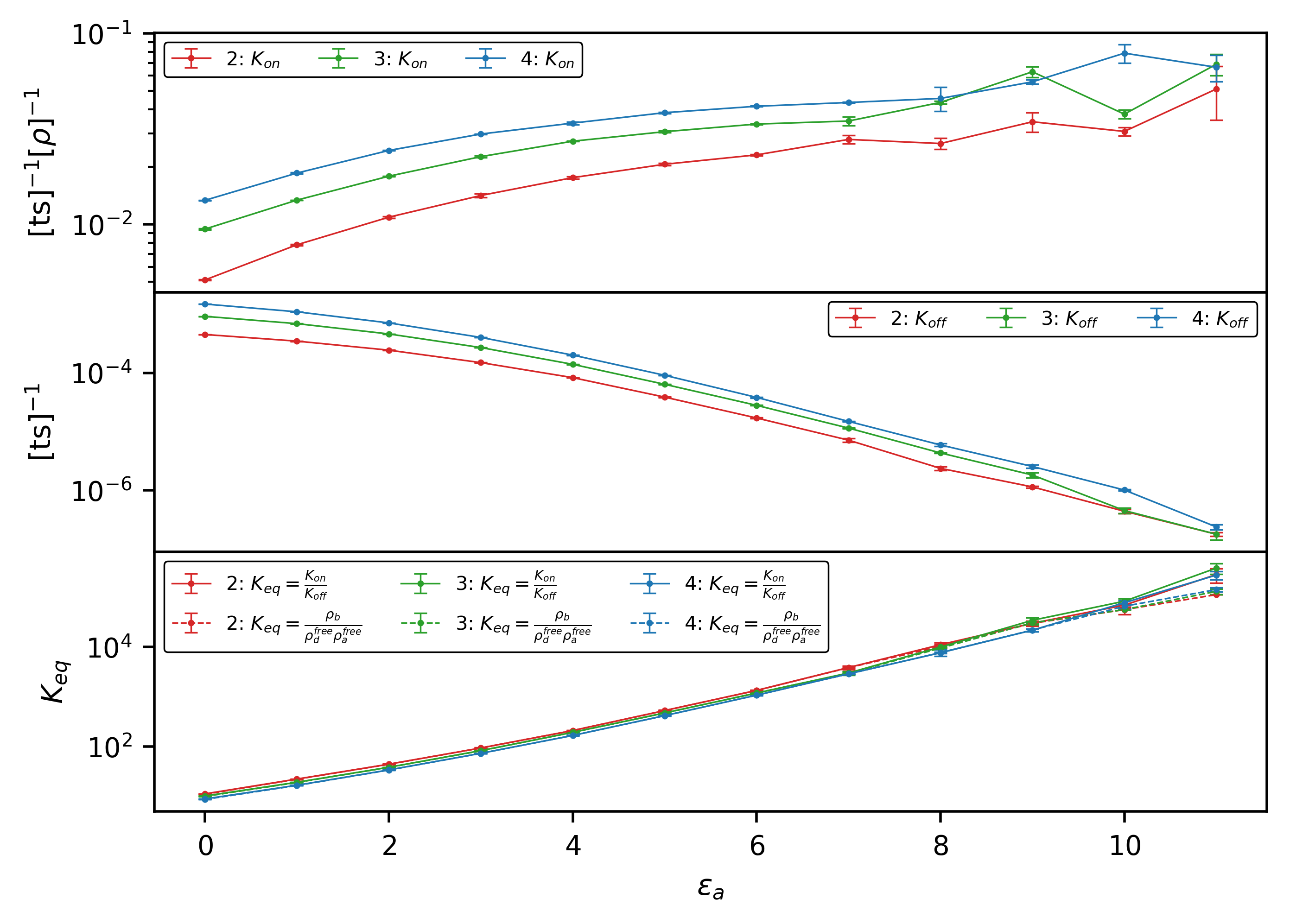}
    \caption{(Top): Rate constants for the bond creation, $k_{\mathrm{on}}$ (solid lines), and bond breaking $k_{\mathrm{off}}$ (dashed lines) as shown in Eq.~\ref{fig:rate}, for the chains with 2 (red), 3 (blue) and 4 (green) active sites; (Bottom): Equilibrium constant of the bond making/breaking reaction $(k_{eq})$ calculated as a ratio of the rate constants $k_{\mathrm{on}}$ and $k_{\mathrm{off}}$ (continuous line), and as a reaction quotient~\ref{eq:keq_rho}.}
    \label{fig:rate}
\end{figure}

Alternatively, $K_{eq}$ can be calculated as
\begin{equation}
 K_{eq}=\frac{\lbrack A-D \rbrack}{\lbrack D_f \rbrack \lbrack A_f \rbrack},
 \label{eq:keq_rho}
\end{equation}
The units of $k_{\mathrm{on}}$ are thus $\frac{1}{\lbrack \rho \rbrack \lbrack \tau \rbrack}$, where $\rho$ is the concentration and $\tau$ denotes time. The units of $k_{\mathrm{off}}$ are just $\frac{1}{\lbrack \tau \rbrack}$.
The rate constants are determined by counting the number of bonds being made and broken over a specified time interval. The exact length of the measurement interval is irrelevant, as long as it is long enough to properly capture the equilibrium properties. Since the concentrations of the free and bound active sites are known at each step, the rates then can be calculated using Eq.~\ref{eq:reaction}. The results are plotted in Figure~\ref{fig:rate}.

As expected, the value of $k_{\mathrm{on}}$ increases with increasing $\varepsilon_a$, whereas $k_{\mathrm{off}}$ decreases as the affinity energy becomes larger. In both cases, the changes are most rapid at low values of $\varepsilon_a$, and become smaller as the active sites become saturated. Whereas large differences in $k_{\mathrm{on}}$ are observed between the systems with different values of $\mathcal{M}$ at the corresponding energies, the effect of $\mathcal{M}$ on $k_{\mathrm{off}}$ is much smaller.
By calculating the value of $K_{eq}$ using two different methods (see equations ~\ref{eq:keq} and~\ref{eq:keq_rho}) we are able to verify the accuracy of our calculations. Similar to $k_{\mathrm{on}}$, the value of $K_{eq}$ increases with increasing $\varepsilon_a$. In agreement with the theoretical predictions of~\cite{danielsen_phase_2023}, we observe exponential scaling of $K_{eq}$ as a function of affinity energy between the active sites. The equilibrium constant of the reaction should only depend on the temperature (and thus on the value of $\varepsilon_a$), and indeed, we observe no variability due to the value of $\mathcal{M}$ itself.

\subsection{Network Structure}
\label{sec:network}

\begin{figure}
    \centering
    \includegraphics[width=.75\textwidth]{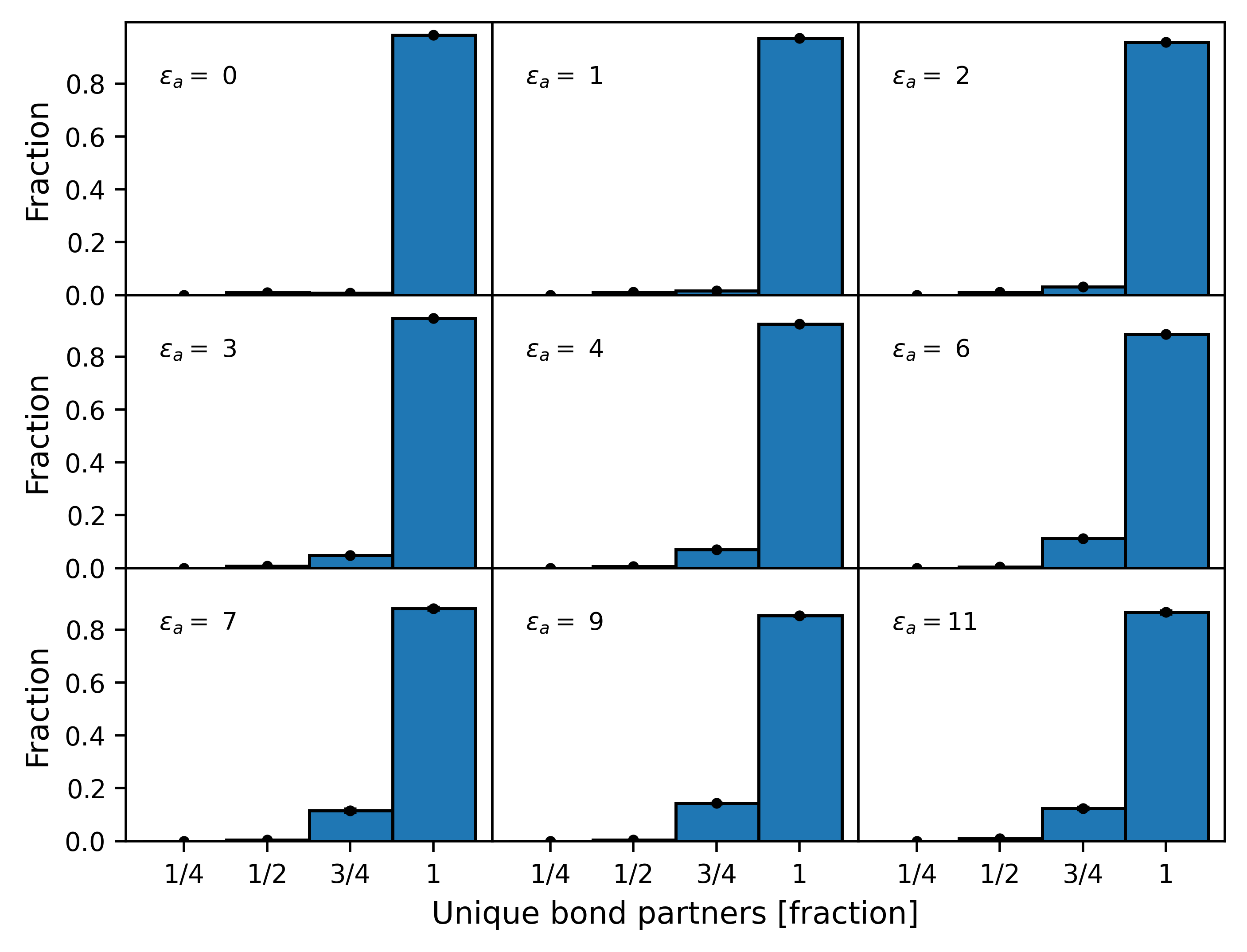}
    \caption{Fraction of unique bonded partners for the system with $\mathcal{M}=4$. The data is calculated by dividing the number of unique bonding partners (separate chains) by the number of bonds a given chain makes. Error bars, indicating one standard deviation, are smaller than the data points (black dots) on top of the histogram bars.}
    \label{fig:Bins2}
\end{figure}

Next, we focus on characterizing the structure of the dynamical network. We begin by looking at the connectivity of the chains. More precisely, we investigate, how many unique binding partners chains tend to have, focusing on the case of $\mathcal{M}=4$. We display the results in Figure~\ref{fig:Bins2}. Two association modes are possible. The chains can either display the preference towards pairwise interactions, and associate with their partner at multiple sites, or they can prefer to attach a different binding partner on each of their sites. In our systems, the second mode (multiple partners) dominates. This behaviour is due to the entropic penalty that two chains would incur if they associated with each other at more than one site, since the mobility and the conformational freedom of these segments would become reduced. Only at the highest values of $\varepsilon_a$ is the affinity energy large enough to compete with the loss of entropy, and the fraction of chains connected to their partner at two sites becomes significant.

\begin{figure}
    \centering
    \includegraphics[width=.75\textwidth]{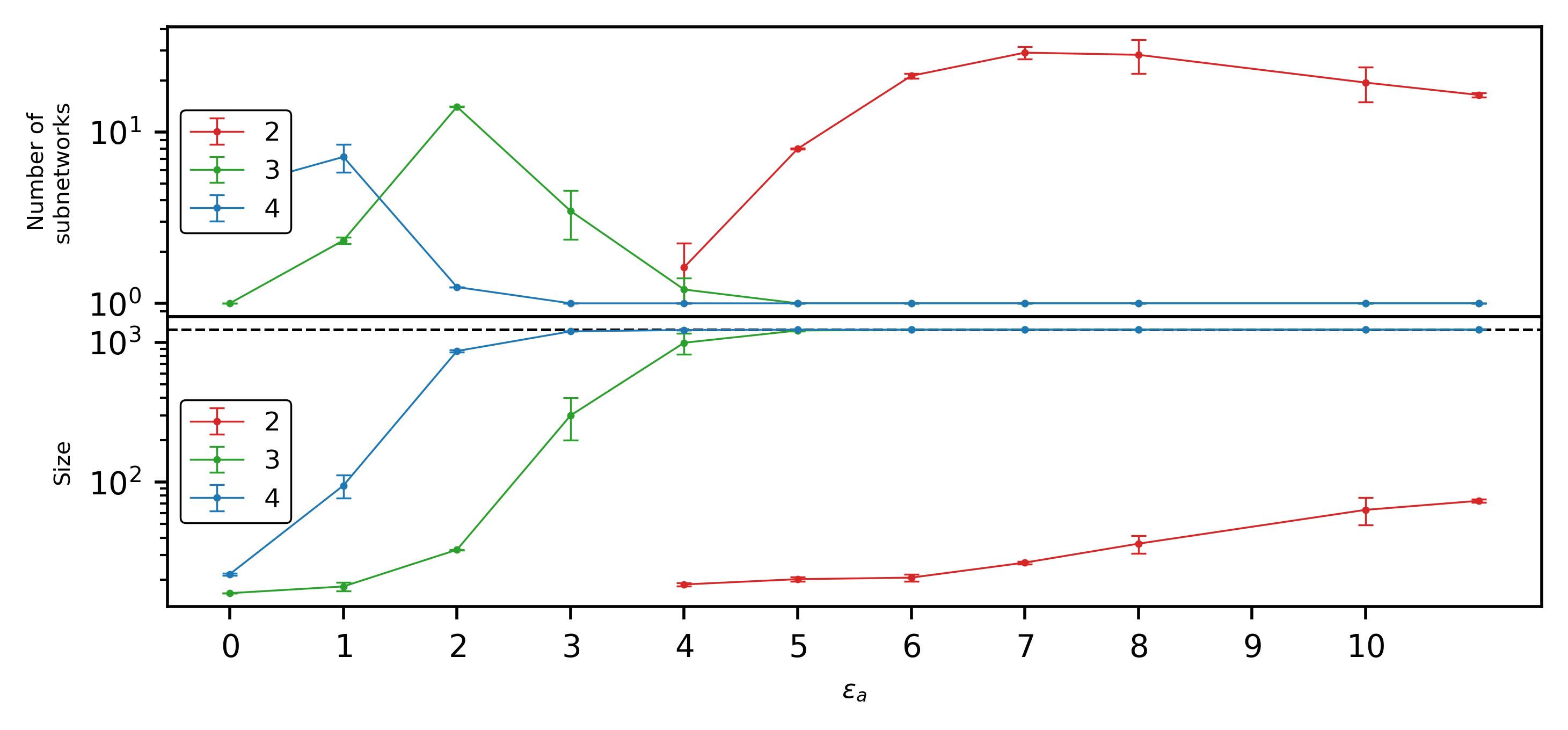}
    \caption{Characterization of the dynamical network. (Top): Average size of a subnetwork defined as a number of chains belonging to the subnetwork. Only subnetworks with size $\geq 15$ chains are included. (Bottom): Fraction of chains bound in subnetworks. The dashed black line corresponds to the total number of chains in the system. Network analysis has been performed using the NetworkX Python package~\cite{NetworkX}.}
    \label{fig:connected-components}
\end{figure}

We further study the network structure by quantifying the size and number of the emergent subnetworks (Figure~\ref{fig:connected-components}). Two chains belong to the same subnetwork if a path created by dynamic bonds exists between them. Here, major differences between the systems with $\mathcal{M}=2$ and $\mathcal{M}\geq3$ are observed. With only two active sites, the chains cannot develop a true network, and instead can only associate as a one, long chain. On the other hand, cross-linking and branching is possible with $\mathcal{M} \geq 3$.

We begin by analyzing the behaviour of systems with $\mathcal{M} = 2$. When the value of $\varepsilon_a$ is low, no ``subnetworks" are observed. As the value of $\varepsilon_a$ increases, the chains begin to assemble into multiple longer chains. At $\varepsilon_a~8$, these chains become to fuse together. This can be seen in the Figure~\ref{fig:connected-components} (red data points), as the number of subnetworks initially increases, reaches the highest points, and then begins to decrease. At the same time, the number of chains being a part of the network continuously increases.

In the case of systems $\mathcal{M}=3$ or 4, subnetworks start to develop even at the lowest values of $\varepsilon_a$ studies. Similar to the case of $\mathcal{M}=2$, the number of subnetworks initially increases, reaches the maximum, and then falls to 1. At the same time, the number of chains which are the part of the subnetworks continuously increases, until it reaches the total number of the chains in the system. At that point all chains belong to a one large network which spans the entire system. The value of $\varepsilon_a$ at which this transition occurs is inversely correlated to the value of $\mathcal{M}$.

Lastly, we asses the presence of ``hoppers'', chains which have released all connections to the network and at low densities can freely diffusive over a large distance \cite{stickers-mechanism}. We quantify the presence of fully unbound ``hoppers'' in our systems by constructing a histogram of the number of occupied active sites on a given chain. The results for the system with $\mathcal{M}=4$ are shown in Figure~\ref{fig:Bins}. We indeed observe a large contribution of free chains at low values of $\varepsilon_a$. The fraction of free chains decreases sharply at the onset of the ``shoulder'' in the density plot, with no free chains present at high values of $\varepsilon_a$. Similar trends are also observed for systems with $\mathcal{M}=2$ and 3.

\subsection{Orthogonal Phase Separation}
\label{sec:phase-sep}

Finally, we investigate the possibility of inducing orthogonal phase separation by introducing a second type of active site. This could most readily be achieved in biological systems with distinct pairs of specific interactions or by synthesizing polymers with multiple mechanisms of association (such as both hydrogen bonding and chelating monomers). Bonds are permitted only between donors and acceptors belonging to the same binding type in our model, without specifying the underlying chemistry. The chains belonging to the same binding type tend to cluster together, which provides a driving force towards orthogonal phase separation. This effect competes with electrostatic forces, which are non-selective and attract oppositely charged donors and acceptors irrespective of their binding type.

To evaluate the possibility of orthogonal phase separation occurring in the systems with varying $\mathcal{M}$, we first initialize systems containing chains belonging to two binding types, homogeneously distributed through the initial slab.
At sufficiently high values of $\varepsilon_a$, partial demixing is observed in systems with $\mathcal{M}=3$ and 4 (data not shown). No demixing is observed when $\mathcal{M}=2$. We do not observe cleanly separated phases, and instead regions consisting predominantly of one binding type are present throughout the coacervate. We attribute it to slow diffusion within these systems, and expect that full demixing will occur at time-scales much longer than those considered here.

\begin{figure}
    \centering
    \includegraphics[width=0.75\textwidth]{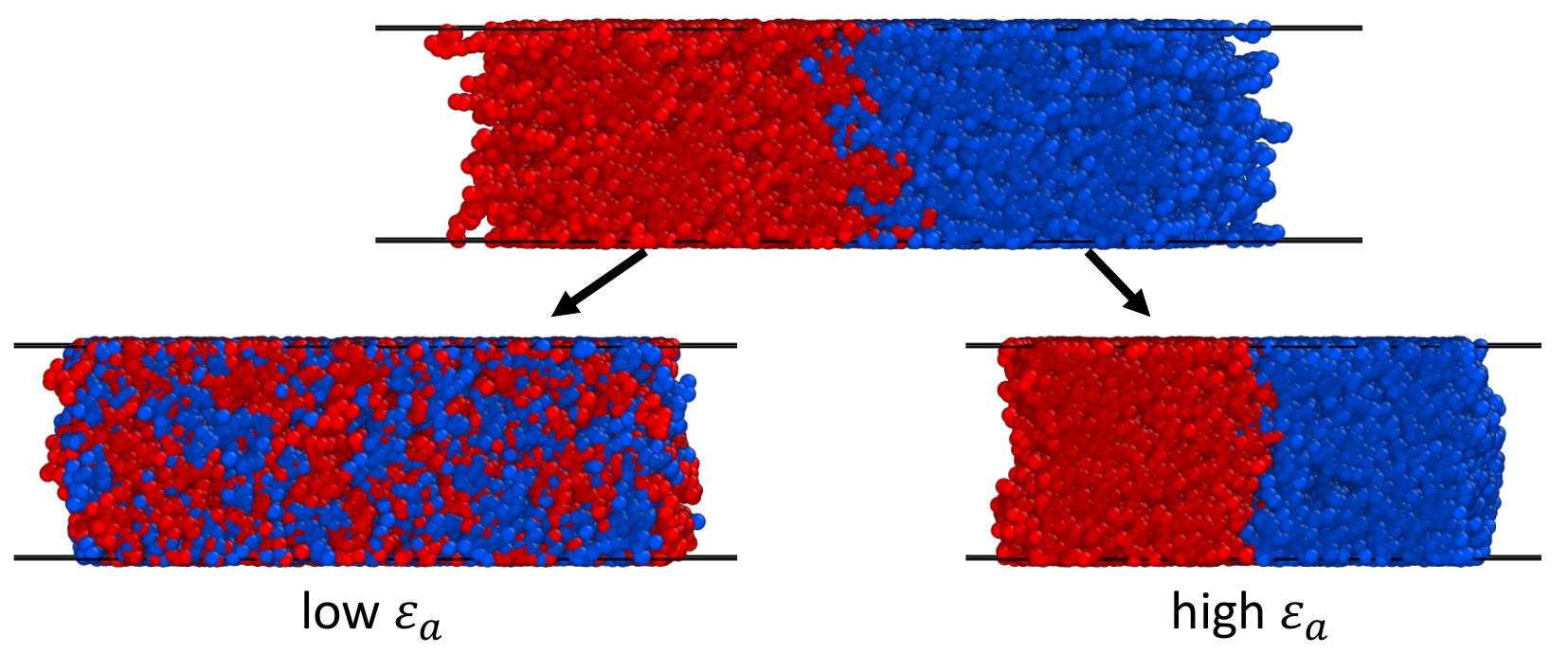}
    \caption{Schematic representation of the orthogonal phase separation in systems with two types of active sites. (Top): Initial configuration of the system with two types types of active sites. Chains belonging to different types are arranged on one side of the box. The chains belonging to the first type are color-coded red, and these belonging to the other type are color-coded blue.
    (Bottom): No phase separation observed at low values of $\epsilon_a$ (left), and orthogonal phase separation at high values of $\epsilon_a$ (right).
    Images were rendered using visualization software Ovito Pro~\cite{stukowski_visualization_2010}.}
    \label{fig:ortgonogonal-phase-sep-combo}
\end{figure}

\begin{figure}
    \centering
    \includegraphics[width=.75\textwidth]{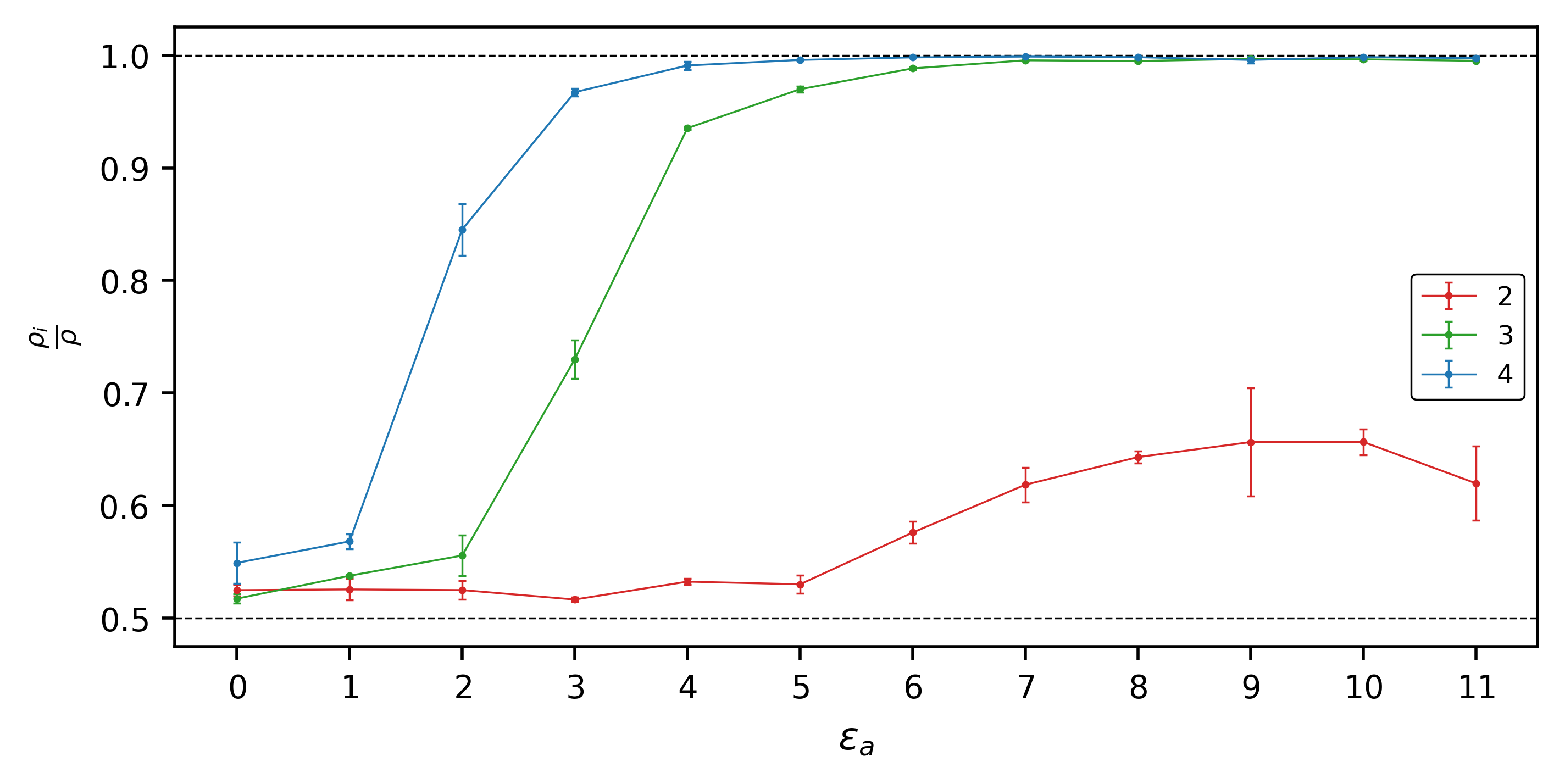}
    \caption{Extent of orthogonal phase separation in the systems with variable number of active sites initialized in a biased configuration. Y-axis corresponds to the ratio of the chains belonging to one of the binding types (on the side of the coacervate they were initialized at) to the total polymer concentration in that region. Demixing persists to a significant extent only in the systems with $\mathcal{M}=3$ and 4.}
    \label{fig:phase-sep-plot}
\end{figure}

In order to quantify the extent of the phase separation, we create new systems with an equimolar ratio of the chains belonging to one of the two binding types. We placed the chains belonging to the first binding type on one side of the box, and the chains belonging to the second binding type are placed on the opposite site (see top panel of Figure~\ref{fig:ortgonogonal-phase-sep-combo}).
We hold these systems at the fixed value of $\epsilon_a$, and allow the chains to diffuse freely across the interface until the systems reach a constant composition. To quantify the degree of the phase separation, we calculate the ratio of the number density of the chains belonging to one phase on their respective side of the interface to the total number density of the polymer on that side.
The results are plotted in the Figure~\ref{fig:phase-sep-plot}. From the data, we see that for the high enough values of $\epsilon_a$, systems with 3 and 4 active sites favor remaining in the demixed state, with the extent of demixing increasing with increasing value of $\epsilon_a$. No demixing persists in the case of $\mathcal{M}=2$ as indicated by a low fraction of the pure phase number density, even at the highest values of bond energy. The value of $\epsilon_a$ required for demixing in the case of $\mathcal{M}=4$ is lower than for $\mathcal{M}=3$. The lack of phase separation in the case of $\mathcal{M}=2$ is likely due to the inability of these chains to create gels, and instead they can only associate at their ends to create long chains. In contrast, chains with $\mathcal{M} \geq 3$ can create highly cross-linked gels.

\section{Conclusions}
\label{sec:conclusions}

In this work we used molecular simulations to investigate the effect of dynamic binding on the properties of charged polymer blends. We studied polymers containing 2, 3 and 4 active sites per chain, separate into donor and acceptor chains. We demonstrated that the phase behaviour of these systems are strongly influenced both by the amount of active sites and variations in the affinity between the donors and acceptors, $\varepsilon_a$. 
All systems began in a polymer-rich coacervate phase driven primary by electrostatics, which transitioned into a higher density polymer phase as the value of $\varepsilon_a$ was increased. We showed that these higher density polymer phases correspond to prominent changes in the the structure of the underlying dynamical network, such as exponential growth in the number of bonds and the development of small subnetworks. We analyzed the differences in diffusive behaviour observed in the systems with varying $\mathcal{M}$. We compared our results with the theoretical predictions of~\cite{Rubinstein-gelation}.
Finally, we studied the possibility of inducing orthogonal phase separation by introducing a second type of active sites into the system. By enforcing that bonds can be created only between the active sites belonging to the same type, we introduce an additional force which favors demixing. We demonstrate that at least 3 active sites per chain are necessary for the orthogonal phase separation to take place. Increasing the value of $\varepsilon_a$ increased the extent of demixing. As orthogonal phase separation is important both in industry and medicine, this observation has the potential to aid the design of materials through self-assembly, and explaining the the processes leading to metabolic disturbance and disease.

\begin{acknowledgement}

This work used Bridges-2 GPU at Pittsburgh Supercomputing Center through allocation DMR150034 from the Advanced Cyberinfrastructure Coordination Ecosystem: Services \& Support (ACCESS) program, which is supported by National Science Foundation grants \#2138259, \#2138286, \#2138307, \#2137603, and \#2138296. 
This work was supported by the National Science Foundation through grant MRSEC-2309043.
This research is part of the Frontera computing project at the Texas Advanced Computing Center. Frontera is made possible by National Science Foundation award OAC-1818253.
The authors acknowledge the Texas Advanced Computing Center (TACC) at The University of Texas at Austin for providing HPC resources (Lonestar6 HPC Cluster) that have contributed to the research results reported within this paper. URL: http://www.tacc.utexas.edu.
Z.M.J is supported by the Frontera Computational Science Fellowships Program, and the Department of Physics and Astronomy at the University of Pennsylvania.
Simulations were performed using MATILDA.FT \cite{jedlinska_matildaft_2023}. Part of the data analysis and generation of images for the publication was done using OVITO Pro \cite{stukowski_visualization_2010}. Plots were generated using the Matplotlib Python module \cite{matplotlib}. Part of the analysis of the network structure was performed using NetworkX Python package \cite{NetworkX}.

\end{acknowledgement}

\section{Appendix A: Algorithm for Dynamic Binding}
\label{sec:appendixA}

Every $\tau_{MC}$ MD steps we perform one MC step. The value of $\tau_{MC}$ has been set to $10,000$ to place us in the reaction-limited regime, which is often observed in experiments \cite{stickers-high-dens}. The reaction-limited regime corresponds to the limit where reaction kinetics are much slower than diffusion of the molecules. In our system, this choice of $\tau_{MC}$ corresponds to the diffusion distance of $\sim 30$ when no dynamic bonds are present but to only $\sim 1$ at the intermediate values of $\varepsilon_a$ and $\mathcal{M}=4$.
Thus, the value of mean-squared displacement when no bonds are present agrees with the solution of a 3D random walk given by
\begin{equation}
 \langle \Delta r^2 \rangle = 6 \tau D_m,   
\end{equation}
where $\Delta r^2$ is the Mean Squared Displacement (MSD), $\tau$ denotes time, and $D_m$ is the monomer diffusivity, which in our system is set to $\frac{1}{75}$.

During each MC step we perform a sequence of binding and unbinding attempts on a fraction of the stickers present in the system. We define the fraction of monomers considered during each two-move sequence as active fraction, $\phi_{active}$, and set $\phi_{active} = 0.05$. We always begin with a binding attempt that is performed on the fraction of particles that were marked as free at the beginning of the sequence, followed by an unbinding attempt that is performed on a fraction of particles that were marked as bound at the beginning of the move sequence (before the bonding move). Since the state of the particles is not updated between bonding and unbinding, the exact choice of the order of moves is immaterial.
After that two-move sequence, we update the lists that keep track of bound and free particles, and particles become marked according to their state. Next, we repeat the two move sequence and state update a certain number of times, which in our case is 2. We verified that varying the value of $\phi_{\mathrm{active}}$ has no effect on the average thermodynamic quantities but keeping this value low helps reduce fluctuations in the system. Since the two-move sequence is repeated at least twice, a particle can in theory be chosen to be active twice, opening a possibility of recombination moves (bond breaking and re-making with the same partner, or switching one bond partner for another).

The probability of accepting a binding or unbinding move on each particle is determined through Metropolis criterion~\cite{MC, metropolis}. First, a random number, $\sigma_r$, is chosen from the range $\lbrack 0,1 \rbrack$. Next it is compared with $P_{\mathrm{on}}$ or $P_{\mathrm{off}}$, which correspond to the probability of making or breaking a bond, respectively. If $\sigma_r < P_{\mathrm{on}}$ or $\sigma_r < P_{\mathrm{off}} $ the move is accepted, otherwise, the move is rejected.
The values of $P_{\mathrm{on}}$ and $P_{\mathrm{off}}$ are determined through setting the desired reaction rates, $k_{\mathrm{on}}$ and $k_{\mathrm{off}}$, with units $\frac{1}{\lbrack \tau \rbrack}$, for binding and unbinding reactions, respectively. The relation between the rate constant and the acceptance probability is given by
\begin{equation}
    P_i = k_i \cdot \tau_{MC},
    \label{eq:Pon}
\end{equation}
where $i$ stands for either ``on" or ``off".
Next, we redefine the ``on" rate as $k_{\mathrm{on}} = k^0_{\mathrm{on}} \phi^s_{\mathrm{on}}$ (with an analogous expression for $k_{\mathrm{off}}$). The rates $k^0_{\mathrm{on}}$ and $k^0_{\mathrm{off}}$ correspond to the spatially-independent energetic part of bonding reaction. Specifically, the affinity energy between the species, $\varepsilon_a$ is given by the ration of these rates
\begin{equation}
\label{eq:epsilon}
    \frac{k^0_{\mathrm{on}}}{k^0_{\mathrm{off}}} = e^{(\varepsilon_a)}.
\end{equation}
The spatially dependent part of $k_{\mathrm{on}}$ and $k_{\mathrm{off}}$ is denoted by $\phi^s_i$, and is responsible for incorporating the spring potential associated with the dynamic bond into the acceptance criterion. Following previous work \cite{marbach_coarse-grained_2023}, we place the spatial dependence both on $k_{\mathrm{on}}$ and $k_{\mathrm{off}}$. In our case, $\phi^s_{\mathrm{on}} = \frac{1}{1+e^{(U_{\mathrm{spring}})}}$ and $\phi^s_{\mathrm{off}} = \frac{1}{1+e^{(-U_{\mathrm{spring}})}}$. Here, $U_{\mathrm{spring}} = \frac{k_{\mathrm{s}}}{2} (r-r_0)^2$ is potential energy of the harmonic bond between two particles separated by distance $r$, with an equilibrium distance of $r_0$, and the spring constant $k_s$.

There are additional details that need to be considered when choosing the value of $k^0_{\mathrm{on}}$. For the correctness of the algorithm, we need to ensure that both $k^0_{\mathrm{on}}$ and $k^0_{\mathrm{off}} \leq 1$ since 1 is the maximum value that probability can have.
Since we can set $t_{\mathrm{on}} = t_{\mathrm{off}} = \tau_{MC}$ to their smallest possible value, we ensure that $k^0_{\mathrm{on}} \leq 1$ by setting $ k^0_{\mathrm{on}} = \frac{e^{(0)}}{t_{\mathrm{on}}} = \frac{1}{\tau_{MC}}$, and then $P_{\mathrm{on}} = \tau_{MC} \cdot k^0_{\mathrm{on}} = 1 $.
This, in turn, determines the choice of $k^0_{\mathrm{off}} = \frac{e^{(-\varepsilon_a)}}{\tau_{MC}}$ (see equation~\ref{eq:epsilon}).

In the following, we explain what happens during each two-move sequence of an MC step.
We begin with the binding phase. We model our approach after~\cite{mitra_coarse-grained_2023} but with certain modifications. First, we generate a neighbour list for each of the donor particles which were free at the beginning of the two-move sequence. This neighbour list contains all nearby acceptor particles, independent of their bonded state. We arrange this from the closest to the furthest particle. Next, we randomly select a donor particle and mark its nearest neighbour to attempt binding. We repeat this action for the remaining donor particles. If the nearest acceptor has already been selected by one of the previous donors, then the next acceptor from the list is chosen, until the list is exhausted. After marking the possible pairs, the binding attempts are performed with the acceptance criterion given by Equation~\ref{eq:Pon}. Our approach differs from that implemented in~\cite{mitra_coarse-grained_2023} since the neighbour list implemented in that work only included the free acceptor particles. We found that when only free acceptor particles were included, the behaviour of the system resembled that of an Ising model~\cite{ising_beitrag_1925} of donor particles, rather than a donor-acceptor system.
Specifically, when only the free acceptors were included in the neighbour list, the bonded fraction at the studied energy have not changed between the systems where donors and acceptors were present in the non-equimolar ratio (data not shown).
Next, during the unbinding phase we simply list all bonded pairs that existed before the binding step, and attempt an unbinding move on the selected fraction of bonded donor particles, $\phi_{\mathrm{active}}$. For each of the selected particles we generate a random number, $\sigma_r$, between 0 and 1. If this number is less or equal to $P_{\mathrm{off}}$ for that specific pair, then the bond is broken; otherwise, it stays intact.

\begin{figure}
    \centering
    \includegraphics[width=0.75\textwidth]{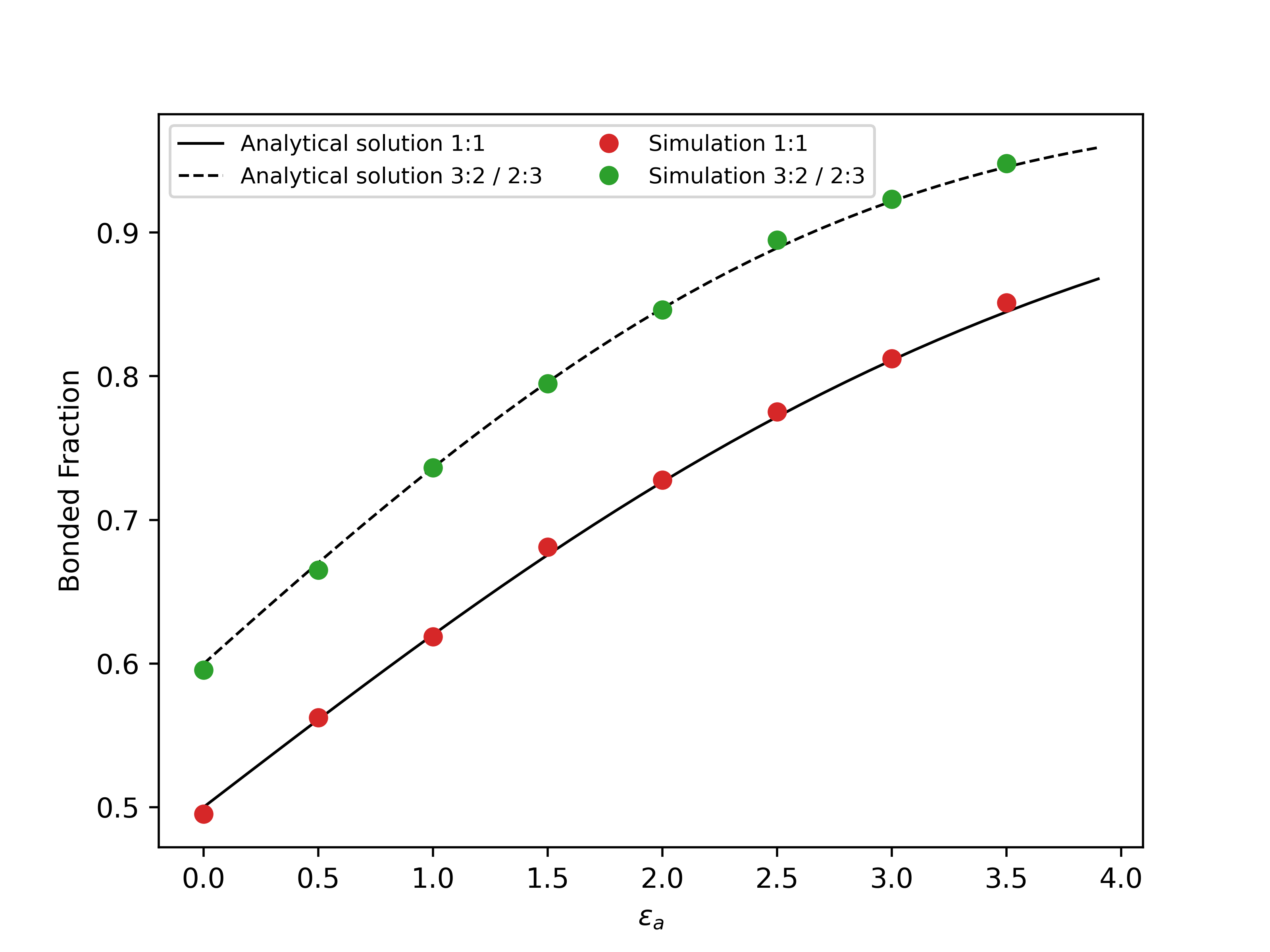}
    \caption{Simulation results, showing the fraction of bonded particles as a function of the affinity energy, $\varepsilon_a$, for the system with an equimolar (red) and non-equimolar (green) number of donor and acceptor particles. Analytical solutions are plotted as black lines, for the equimolar (dashed) and non-equimolar (continuous) systems.}
    \label{fig:Figure1A}
\end{figure}

Below, in Figure \ref{fig:Figure1A}, we present a verification of our algorithm by comparing the results with an analytical solution, the derivation of which is provided below.

We verify that the base algorithm obeys detailed balance by comparing the results of our simulations with the analytical solution for a test system. The test system consists of a mixture of $n_D$ donors particles (D) and $n_A$ acceptor particles (A). Any D particle can form at most one bond with an A particle, such that at any given time a particle cannot have more than one binding partner, and only D-A pairs are permitted. There is no spatial dependence ($k_{spring} = 0.0$), so that the equilibrium bonded fraction depends only on the energetic difference between the bound and unbound states, $\varepsilon_a$, and the entropic contribution arising from the number of ways for choosing the identities of the pairs involved.
The reaction is given by
\begin{equation}
 \ce{ $D + A$ \leftrightharpoons $B$ },
\end{equation}
where $B$ denotes a bonded pair. When a bond is formed, the energy of the system is lowered by $\varepsilon_a$, giving the partition function
\begin{equation}
    Z = \sum_{n_B = 0}^{n_{max}}\Omega(n_B)e^{(n_B \cdot \varepsilon_a)} 
\end{equation}
where the summation goes over the number of bonded pairs in the system. Here, $n_{max}$ is the maximum possible number of bonds in the system, equal to $\mathrm{min}(n_A,n_D)$, and $\Omega(n_B)$ is the density of states corresponding to the given number of bonded pairs, $n_B$; i.e., $\Omega(n_B)$ is the number of indistinguishable ways of creating $n_B$ bonded pairs. We can calculate $\Omega(n_B)$ by noting that without spatial dependence it is just the number of ways to arrange $n_B$ donors and acceptors in pairs, given by
\begin{equation}
    \Omega(n_B) = \frac{n_A!}{n_B!(n_A - n_B)!} \cdot \frac{n_D!}{n_B!(n_D - n_B)!} 
\end{equation}
The probability of finding a given number of bonded pairs is then given by 
\begin{equation}
    P(n_B) = \frac{\Omega(n_B)e^{(n_B \cdot \varepsilon_a)}}{Z}
\end{equation}
The result of a simulation with 1,000 particles and variable donor to acceptor ratio is shown in Figure~\ref{fig:Figure1A} along with the corresponding analytical solutions.

\begin{figure}
    \centering
    \includegraphics[width=0.75\textwidth]{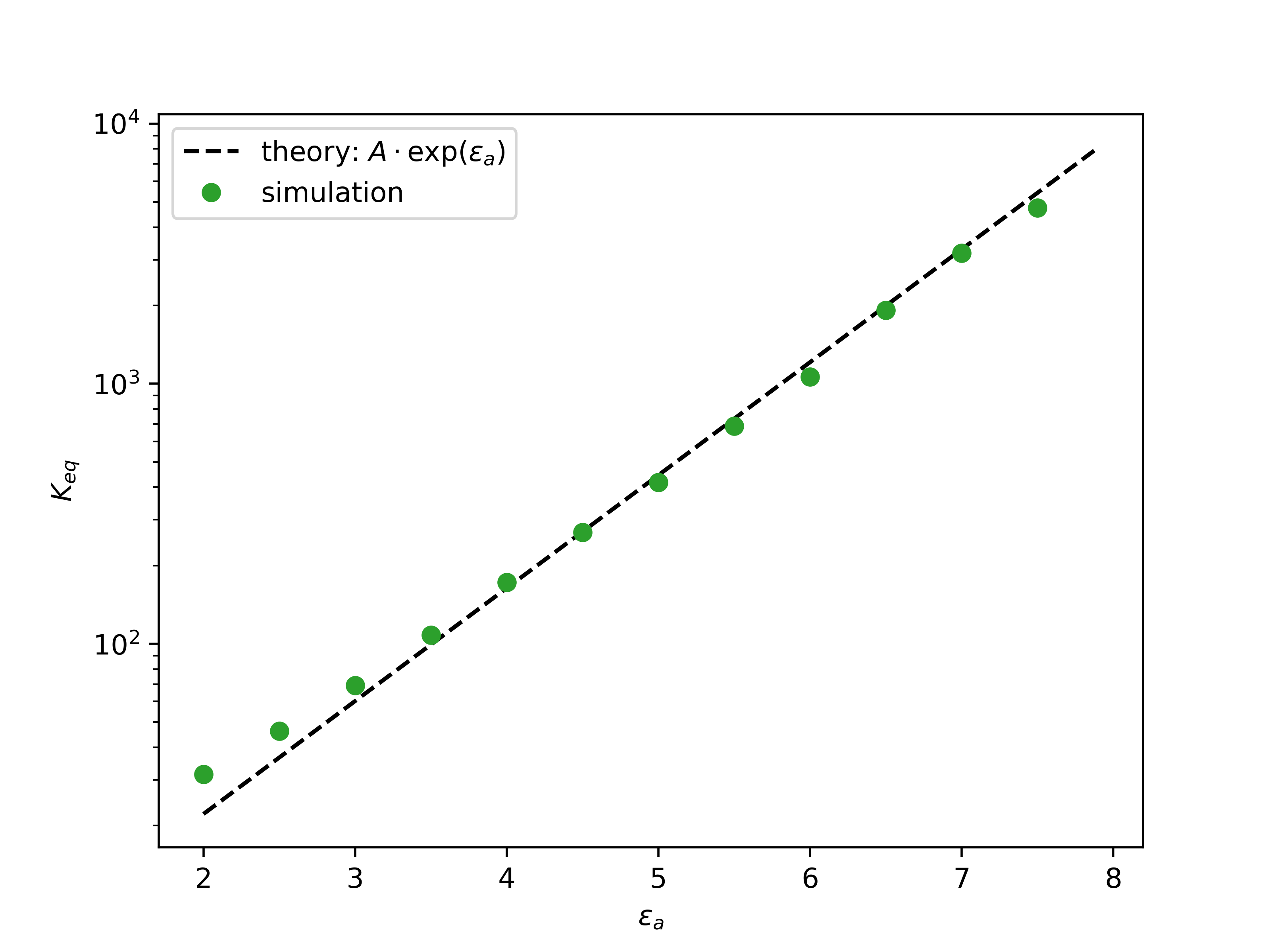}
    \caption{Scaling of the equilibrium constant, $K_{eq}$, as a function of $\varepsilon_a$. Results of the simulation are shown as green data points, while the theoretically predicted scaling is shown as a black, dashed line.}
    \label{fig:Figure2A}
\end{figure}

We validate that the algorithm behaves as expected when $k_{spring} \neq 0.0$ by plotting the equilibrium constant, $K_{eq}$, against the bond formation energy, $\varepsilon_a$. $K_{eq}$ is given by
\begin{equation}
    K_{eq} = \frac{\lbrack n_B \rbrack}{\lbrack n_D^f \rbrack \lbrack n_A^f \rbrack}
\end{equation}
where the square bracket denote concentrations and $\lbrack n_D^f \rbrack$ and $\lbrack n_D^f \rbrack$ denote the concentration of free donors and acceptors respectively. As expected \cite{fredrickson_2009, danielsen_phase_2023}, we observe that $K_{eq} \sim e^{(\varepsilon_a)}$. Results of the simulation along with an exponential fit are shown in Figure~\ref{fig:Figure2A}.

\bibliography{bibliography}

\end{document}